\documentclass[showpacs, amsmath,amssymb, aps]{revtex4}

\usepackage{tabularx}
\usepackage{enumitem}
\usepackage{graphicx}
\usepackage{float}
\usepackage{dcolumn}
\usepackage{bm}
\usepackage{mathrsfs}
\usepackage{amsmath}

\newcommand{\Splus}{S_{+\frac{1}{2}}}
\newcommand{\Sminus}{S_{-\frac{1}{2}}}
\usepackage{amsfonts}

\begin{document}

\title{On the eigenvalues of the fermionic angular eigenfunctions in the Kerr metric}
\author{D. Batic}
\email{davide.batic@ku.ac.ae}
\affiliation{%
Department of Mathematics,\\  Khalifa University of Science and Technology,\\ Main Campus, Abu Dhabi,\\ United Arab Emirates}
\author{S. Hamad Abdul Karim}
\email{100049756@ku.ac.ae}
\affiliation{%
Department of Mathematics,\\  Khalifa University of Science and Technology,\\ Main Campus, Abu Dhabi,\\ United Arab Emirates}
\author{M. Nowakowski}
\email{mnowakos@uniandes.edu.co}
\affiliation{
Departamento de Fisica,\\ Universidad de los Andes, Cra.1E
No.18A-10, Bogota, Colombia
}

\date{\today}

\begin{abstract}
In view of a result recently published in the context of deformation theory of linear Hamiltonian systems, we reconsider the eigenvalue problem associated to the angular equation arising after the separation of the Dirac equation in the Kerr metric and we show how efficiently a quasi-linear first order PDE for the angular eigenvalues can be derived. We also prove that it is not possible to obtain an ordinary differential equation for the eigenvalues where the role of the independent variable is played by the particle energy or the black hole mass. Finally, we construct new perturbative expansions for the eigenvalues in the Kerr case and obtain an asymptotic formula for the eigenvalues in the case of a Kerr naked singularity.
    
\end{abstract}
\pacs{xyz}
\maketitle
\section{Introduction}
The Chandrasekhar-Page (CP) equations describe the angular part of the wave function of a spin$1/2$ particle in the presence of a Kerr or Kerr-Newman black hole (BH). More precisely, they emerge from the search for a separable solution to the Dirac equation in the aforementioned geometries. This was achieved by Chandrasekhar \cite{Chandra} in 1976  and extended the same year by \cite{Page} to the case of rotating charged BHs. The Kerr-Newman BH, even though it represents a generalization of spinning BHs, it cannot be considered a realistic astrophysical object because in Nature we generally observe BHs whose surroundings are filled with gas and plasma that would have the effect of discharging the BH.

Since the seminal work of \cite{Chandra,Page} there has been a steady growth of interest in studying the interaction of fermions with rotating BHs probably due to the fact that in this context the Dirac equation couples the BH rotation parameter with the quantum mechanical spin of the particle and the BH mass with the particle mass. Far from giving a complete review on this topic, let us recall that an integral representation for the Dirac propagator was derived by means of two different approaches in \cite{Finster,Batic}. Such a representation is at the core of the time-dependent scattering theory for massive Dirac particles developed in \cite{Batic1}. For other studies centred on the scattering of massive, spin-half particles from a spinning BH performed by numerical methods and a WKB approximation, we refer to \cite{muko1} and \cite{muko2}, respectively. Local energy decay of solutions of the Dirac equation in the non-extreme Kerr metric leading to the abence of bound state solutions was studied by \cite{Yam}. The problem of quantizing fermions on a Kerr manifold, also known as kermions, was studied by \cite{Casals}. Moreover, \cite{Dolan2015} analyzed the gravitationally-trapped modes (regular across the future event horizon) for a massive neutral spin-half particle on a rotating BH spacetime while \cite{rum} showed how the generalized Heun equations governing the radial and angular parts of the Dirac equation can be reduced to the confluent Heun case. 

The angular part of the Dirac equation has been extensively studied. For instance, \cite{Suff} constructed expansions of the angular eigenfunctions in terms of hypergeometric polynomials and derived a transcendental equation for the eigenvalues while  \cite{Kalnins} obtained series representations of the angular eigenfunctions involving Jacobi polynomials and calculated the spectrum of eigenvalues by means of continued fraction methods. Moreover, \cite{Batic-Schmid05} optimized the findings in \cite{Suff,Kalnins} by determining a general recurrence relation for the coefficients entering in a certain power series expansion for the eigenvalues and showed that the eigenvalues fulfil a first order partial differential equation with respect to the physical parameters of the problem. For a variational principle for block operator matrices applied to the angular part of the Dirac operator in the presence of a Kerr BH, we refer to \cite{Monika}. It is worth mentioning that the authors in \cite{Dolan2009} not only uncovered and corrected some errors in the literature but they also developed a new technique for constructing solutions of the CP equation based on a representation for the spin-half spherical harmonics first developed in \cite{Berti}. Finally, a recent effective method to calculate the eigenvalues of the CP equations making use of the Pr\"{u}fer transformation has been proposed by \cite{russ}.

In the present work, we focus instead on the angular part of the Dirac operator in a Kerr BH. The results we obtain are also valid in the case of an extreme Kerr BH upon identification of the spinning parameter with the total mass of the black hole. For a discussion of the relevance of such BHs in astrophysics, their connection to relativistic Dyson rings and their emergence as the only possible candidate for a black hole limit for stationary and axisymmetric, uniformly rotating perfect fluid bodies with a cold equation of state as well as for isentropic stellar models with a non-zero temperature, we refer to \cite{5,6,7,8,9,10}. The interaction of fermions with this kind of BHs has been studied in several circumstances. For instance, \cite{Schmid} suggested the existence of bound state solutions for the Dirac equation coupled to the extreme Kerr metric. \cite{BS} derived an integral representation of the Dirac propagator in that metric. Moreover, the absence of stationary bound states of half-spin particles with extreme Kerr and Kerr-Newman backgrounds was proved by \cite{Uns} and \cite{Rus1}, respectively. The study of Dirac particles in the presence of an extreme Kerr BH despite the fact of being a well-established research field still hides some surprises. To be more precise, we recall that by a fortuitous method outlined in \cite{Batic-Schmid05}, it was possible to derive there a first order quasi-linear partial differential equation (PDE)  for the  eigenvalues of the angular equation arising after separation of variables is applied to the Dirac equation in the geometry of a spinning black hole. At that time, it was not clear whether this approach was limited to that particular problem or not. Sixteen years later, \cite{Schmid21} proved a theorem stating under which general conditions the eigenvalues of a linear Hamiltonian system satisfy a first order quasi-linear partial differential equation with respect to the relevant physical parameters. Inspired by this result, we checked through an explicit computation of the deformation matrix that the PDE obtained in \cite{Batic-Schmid05} for a Kerr black hole can indeed be derived from the general theory developed in \cite{Schmid21}. However, we did not only limit ourselves to this exercise. We also showed that in the context of the deformation theory of linear Hamiltonian systems a simple ODE for the angular eigenvalues can be efficiently derived whose solution agrees with that given by \cite{Batic-Schmid05}.

We believe that the method described here will not only be useful for the study of the emission of massive fermions from rapidly spinning BHs in higher dimensions since the angular equations of Dirac particles on the brane are strongly linked to the corresponding four-dimensional counterparts \cite{DOL} but also for a variety of other  boundary value problems arising in physics and engineering. The rest of our work is organized as follows. In Section~\ref{LHS}, we introduce the basic equations and results of the theory of linear Hamiltonian systems needed in the reminder of the paper. In particular, we show how the PDE for the eigenvalues of the angular problem in the Kerr metric found in \cite{Batic-Schmid05} can be efficiently derived in the framework of the aforementioned theory by offering an explicit computation of the deformation matrix. As a by-product, we are able to show that only for a certain relation between the independent variables it is possible to have an ODE for the eigenvalues and we verify that its solution agrees with equation $(45)$ in \cite{Batic-Schmid05} and a similar formula in \cite{Suff}. We also prove that it is impossible to obtain an ODE for the eigenvalues where the independent variable is the particle energy or the BH mass. Section~\ref{EIGENODE} is devoted to the derivation of new perturbative expansions for the eigenvalues. A nice feature of this result is that the expansion coefficients are not affected by the problem of dividing by numbers which may be zero as instead it was the case for certain power series representations of the eigenvalues obtained by \cite{Suff,Kalnins,Batic-Schmid05}. The main difference between our approach and that in the aforementioned literature is that we work with a PDE which is directly formulated in terms of the physical variables of the problem, i.e. the BH spinning parameter and the particle energy. Finally, under the assumption that the BH spin exceeds the total mass of the BH, we also obtain a new asymptotic formula for the angular eigenvalues in the case of a naked rotating singularity. In Section~\ref{Conclusions} we draw our conclusions.

\section{Linear Hamiltonian systems and the CP equations}\label{LHS}
\cite{Schmid21} showed that the derivation of a PDE for the eigenvalues of the CP equations was not a lucky stroke because similar PDEs can also be obtained for other differential operators provided that some technical conditions are satisfied. In the following, we briefly present the main result in \cite{Schmid21} needed for the reminder of our work. By $\lambda$ and $E_n$ we denote the spectral parameter and the $n\times n$ identity matrix, respectively. A linear Hamiltonian $2n\times 2n$ system is represented by the equation
\begin{equation}\label{1}
J\frac{dy}{dx}=\left[\lambda W(x)+H(x;p_1,\cdots,p_m)\right]y,\quad
J=\left(\begin{array}{cc}
0 & E_n\\
-E_n & 0
\end{array}\right)
\end{equation}
with the coefficient matrix $H$ exhibiting a dependence on the independent variable $x$ and on $m$ parameters $p_i$ while the coefficient matrix $W$ also known as the weight function depends on $x$ only. In addition, we require that $x\in(a,b)\subset\mathbb{R}$ with $a<b$ and $(p_1,\cdots,p_m)\in U\subset\mathbb{R}^m$ where $U$ is a non-empty connected open subset. Let us suppose that
\begin{enumerate}
\item
$W:(a,b)\longrightarrow M_{2n}(\mathbb{C})$ and $H:(a,b)\times U\longrightarrow M_{2n}(\mathbb{C})$ are at least twice continuously differentiable. Moreover, both $W$ and $H$ are Hermitian matrices, i.e. $W^\dagger(x)=W(x)>0$ for all $x\in(a,b)$ and $H^\dagger=H$ for all $(x;p_1,\cdots,p_m)\in(a,b)\times U$.
\item
For all $(p_1,\cdots,p_m)\times U$ the operator $T=T(p_1,\cdots,p_m)$ is a self-adjoint extension of the differential operator $\tau$ defined as
\begin{equation}\label{tau}
\tau y:=W^{-1}\left(J\frac{dy}{dx}-Hy\right),\quad D(T)\subset D(T_{max}),
\end{equation}
where $D(T_{max})$ denotes the domain of definition of the maximal operator $T_{max}y:=\tau y$ with
\begin{equation}
D(T_{max})=\left\{y\in L^2_W\left((a,b),\mathbb{C}^{2n}\right)~|~y\in AC_{loc}((a,b),\mathbb{C}^{2n}),~\tau y\in L^2_W\left((a,b),\mathbb{C}^{2n}\right)\right\}.
\end{equation}
Here, $AC_{loc}(X,Y)$ denotes the space of all locally absolutely continuous functions $y:X\longrightarrow Y$ and 
\begin{equation}
L^2_W\left((a,b),\mathbb{C}^{2n}\right)=\{f:(a,b)\longrightarrow\mathbb{C}^{2n}~|~\int_a^b f^\dagger(x) W(x) f(x)~dx<\infty\}.
\end{equation}
\item
For all $(p_1,\cdots,p_m)\times U$ $\lambda(p_1,\cdots,p_m)$ is an eigenvalue of $T=T(p_1,\cdots,p_m)$ and $y(x;p_1,\cdots,p_m)$ is the associated normalized eigenfunction, that is
\begin{equation}
\langle y,y\rangle=\int_a^b y^\dagger(x)W(x)y(x)~dx=1
\end{equation}
with $y:(a,b)\times U\longrightarrow\mathbb{C}^{2n}$ at least twice continuously differentiable.
\item
Let $G:(a,b)\times\mathbb{C}\times\mathbb{C}^m\longrightarrow M_{2n}(\mathbb{C})$ be a differentiable matrix function and $f_k,g:\mathbb{C}\times\mathbb{C}^m\longrightarrow\mathbb{C}$ with $k=1,\cdots,m$ be such that $G=G(x;\lambda;p_1,\cdots,p_m)$, $f_k=f_k(\lambda;p_1,\cdots,p_m)$ and $g=g(\lambda;p_1,\cdots,p_m)$ fulfil the deformation equation
\begin{equation}\label{defeq}
\frac{\partial G}{\partial x}+\left(\lambda W+H\right)JG-GJ(\lambda W+H)=\sum_{k=1}^m f_k\frac{\partial H}{\partial p_k}+gW.
\end{equation}
\end{enumerate}
Then, \cite{Schmid21} proved that if in addition (\ref{tau}) is in the limit point case at $a$ and $b$ and
\begin{equation}
\sum_{k=1}^m f_k\frac{\partial y}{\partial p_k}+JGy\in L^2_W\left((a,b),\mathbb{C}^{2n}\right)
\end{equation}
for all $(p_1,\cdots,p_m)\in U$, the eigenvalues $\lambda$ satisfy the quasi-linear PDE
\begin{equation}\label{PDEeig}
\sum_{k=1}^m f_k\frac{\partial\lambda}{\partial p_k}=g(\lambda;p_1,\cdots,p_m).
\end{equation}
In the case the Hamiltonian system depends linearly on the parameters, that is
\begin{equation}
H(x;p_1,\cdots,p_m)=H_0(x)+\sum_{k=1}^m p_k H_k(x),
\end{equation}
\cite{Schmid21} showed that some of the conditions listed above admit an equivalent formulation that is easier to be verified. While conditions $1.$ and $4.$ remain unchanged, conditions $2.$ and $3.$ can be cast into the following ones
\begin{enumerate}[label=(\roman*)]
\item
The differential expression
\begin{equation}\label{tau0}
\tau_0 y:=W^{-1}\left(J\frac{dy}{dx}-H_0 y\right)
\end{equation}
is singular on $(a,b)$ and in the limit point case at both boundary points. This ensures that the closure of the associated minimal operator $T_0=T_{min}$ is the only self-adjoint extension of $\tau_0$ in $L^2_W\left((a,b),\mathbb{C}^{2n}\right)$.
\item 
$\lambda_0$ is a simple real eigenvalue of $T_0$ and $\lambda$ is an isolated eigenvalue in the spectrum of $T_0$ with algebraic multiplicity one. 
\end{enumerate}
Since \cite{Schmid21} did not provide an explicit derivation of the PDE governing the eigenvalues of the CP equations, we fill this gap in the second part of this section. Let us recall that in Boyer-Lindquist coordinates $(t,r,\vartheta,\varphi)$ such that $r>0$, $\vartheta\in[0,\pi]$ and  $0\varphi\in[0,2\pi)$ the Kerr is expressed by means of the line element \cite{Wald}
\begin{equation}
ds^{2}=\frac{\Delta-a^{2}\sin^{2}{\vartheta}}{\Sigma}dt^{2}
+\frac{2a\sin^{2}{\vartheta}(r^2+a^2-\Delta)}{\Sigma}dt d\varphi
-(r^2+a^2)^{2}\,\sin^2{\vartheta} \frac{\widetilde{\Sigma}}{\Sigma}d\varphi^{2}
-\frac{\Sigma}{\Delta}dr^2-\Sigma d\vartheta^{2},
\end{equation}
where
\begin{equation}
\Sigma:=r^2+a^2\cos^{2}\theta,\quad \Delta:=r^2-2Mr+a^2+Q^2,\quad
\widetilde{\Sigma}:=1-\frac{a^2\Delta}{(r^2+a^2)^2}\sin^2{\vartheta}.
\end{equation}
Here $M$ and $a$ are the mass and the angular momentum per unit mass of the BH. The case of a non-extreme Kerr BH is represented by the inequality $a^2<M^2$ which corresponds to the condition for the function $\Delta$ to have two distinct real roots located at
\begin{equation}
r_{\pm}:=M\pm\sqrt{M^2-a^2}.
\end{equation}
We recall that $r_{-}$ corresponds to the Cauchy horizon while $r_+$ is the BH event horizon. In the extreme case corresponding to $M^2=a^2$ the aforementioned horizons coalesce to form a double root for $\Delta$ at $r_{+}=M=r_{-}$. Finally, there is a ring-like naked singularity whenever $a^2>M^2$. Moreover, the Dirac equation coupled to a general gravitational field can be expressed in terms of two-component spinors $(\phi^{A},\chi^{A^{'}})$ by \cite{pen}
\begin{equation}
\nabla^{A}_{A^{'}}\phi_{A}=\frac{m}{\sqrt{2}}\chi_{A^{'}},\qquad   
\nabla_{A}^{A^{'}}\chi_{A^{'}}=\frac{m}{\sqrt{2}}\phi_{A}
\end{equation}
where Planck units $\hbar=c=G=1$ have been adopted. Note that $\nabla_{AA^{'}}$ is the symbol for covariant differentiation and $m$ is the Fermion rest mass. As already mentioned before, the Dirac equation in the Kerr geometry was computed and separated by \cite{Chandra,Page} using the Kinnersley tetrad \cite{kin}. However, it is more convenient to achieve the separation with the Carter tetrad \cite{Carter} together with the ansatz \cite{Chandra,KN}
\begin{equation}
\Psi:=\left( \begin{array}{c}
               -\phi_{0}\\
               +\phi_{1}\\
               i\chi_{1^{'}}\\
               i\chi_{0^{'}}
\end{array}\right)=e^{i\omega t}e^{ik\varphi}S(r,\vartheta)\left( \begin{array}{c}
               R_{-\frac{1}{2}}(r)S_{-\frac{1}{2}}(\vartheta)\\
               R_{+\frac{1}{2}}(r)S_{+\frac{1}{2}}(\vartheta)\\
               R_{+\frac{1}{2}}(r)S_{-\frac{1}{2}}(\vartheta)\\
               R_{-\frac{1}{2}}(r)S_{+\frac{1}{2}}(\vartheta)
\end{array} \right),\quad k=\kappa+\frac{1}{2},
\end{equation}
where $\omega$ is the energy of the particle as measured at infinity, $\kappa$ is a half-integer such that $|\kappa|\geq 1/2$ with $\kappa=k-1/2$ and  $k\in\mathbb{Z}$. Moreover, $S$ denotes the non singular matrix 
\begin{equation}
S(r,\vartheta):=\Delta^{-\frac{1}{4}}~\mbox{diag}\left(\frac{1}{\sqrt{r-i a\cos{\vartheta}}},\frac{1}{\sqrt{r-i a\cos{\vartheta}}}, \frac{1}{\sqrt{r+i a\cos{\vartheta}}},\frac{1}{\sqrt{r+i a\cos{\vartheta}}}\right).
\end{equation}
Then, the Dirac equation decouples into the following systems of linear first order differential equations for the radial $R_{\pm\frac{1}{2}}$ and angular components $S_{\pm\frac{1}{2}}$ of the spinor $\Psi$ 
\begin{equation} 
\left( \begin{array}{cc}
     \sqrt{\Delta}\mathcal{D}^{-}_\frac{1}{2}&-i mr-\lambda\\
     i mr-\lambda&\sqrt{\Delta}\mathcal{D}^{+}_\frac{1}{2}
           \end{array} \right)\left( \begin{array}{cc}
                                     R_{-} \\
                                     R_{+}
                                     \end{array}\right)=0,\quad
\left( \begin{array}{cc}
     -\mathcal{L}^{-}_\frac{1}{2} & \lambda+am\cos{\vartheta}\\
                \lambda-am\cos{\vartheta} & \mathcal{L}^{+}_\frac{1}{2}
           \end{array} \right)\left( \begin{array}{cc}
                                     S_{-\frac{1}{2}} \\
                                     S_{+\frac{1}{2}}
                                     \end{array}\right)=0 \label{angular}
\end{equation} 
with
\begin{eqnarray}
&&\mathcal{D}^{\pm}_\frac{1}{2}:=\frac{d}{dr}\mp i\frac{K(r)}{\Delta},\quad K(r)=\omega(r^2+a^2)+ak, \label{ramb1}\\
&&\mathcal{L}^{\pm}_\frac{1}{2}:=\frac{d}{d\vartheta}+\frac{1}{2}\cot{\vartheta}\pm Q(\vartheta),\quad Q(\vartheta)=a\omega\sin{\vartheta}+\kappa\csc{\vartheta} \label{ramb2}.
\end{eqnarray}
Finally, the CP equations for a fermion in the Kerr metric are \cite{Batic-Schmid05}
\begin{eqnarray}
\mathcal{L}_{\frac{1}{2}}^{+}\Splus&=&(\mu\cos{\vartheta}-\lambda)\Sminus,\label{CPAE1}\\
\mathcal{L}_{\frac{1}{2}}^{-}\Sminus&=&(\mu\cos{\vartheta}+\lambda)\Splus,\label{CPAE2}
\end{eqnarray}
with $\mu:=am$ and $\nu:=a\omega$. By means of the transformations $x=\sin^2{\vartheta/2}$ mapping the interval $(0,\pi)$ to the interval $(0,1)$ and 
\begin{equation}\label{STrans1}
\begin{pmatrix}
        {\Splus(\vartheta)}\\
        {\Sminus(\vartheta)}
\end{pmatrix}:=\begin{pmatrix}
        \sec{\frac{\vartheta}{2}} & 0\\
        0 & \csc{\frac{\vartheta}{2}}
    \end{pmatrix} y(x),\quad
y(x):=\begin{pmatrix}
        y_+(x)\\
        y_{-}(x)
\end{pmatrix}
\end{equation}
we can cast (\ref{CPAE1}) and (\ref{CPAE2}) into the matrix form
\begin{equation}\label{intra}
    \frac{dy}{dx}=\left(\frac{B_0}{x}+\frac{B_1}{x-1}+C\right)y(x)
\end{equation}
with coefficient matrices
\begin{equation}
B_0:=\begin{pmatrix}
    -\frac{\kappa}{2}-\frac{1}{4} & \mu-\lambda\\
    0 & \frac{\kappa}{2}+\frac{1}{4}
    \end{pmatrix},\quad 
B_1:= \begin{pmatrix}
    \frac{\kappa}{2}+\frac{1}{4} & 0 \\
    \mu-\lambda & -\frac{\kappa}{2}-\frac{1}{4}
    \end{pmatrix},\quad 
C:= \begin{pmatrix}
    -2\nu & -2\mu \\
    2\mu & 2\nu
    \end{pmatrix}.
\end{equation}
At this point, if we multiply (\ref{intra}) by the matrix $J$ defined in (\ref{1}) and rearrange terms accordingly, we can easily bring (\ref{intra}) into the form of the Hamiltonian system (\ref{1}) with $n=1$. More precisely, we find
\begin{equation} \label{CPAEHamil}
    J\frac{dy}{dx} = \left[\lambda W+H(x;\mu,\nu)\right]y(x),\quad
 W=\begin{pmatrix}
\frac{1}{1-x} & 0\\
0 & \frac{1}{x}
\end{pmatrix},\quad
H(x;\mu,\nu)=H_0(x)+\mu H_1(x)+\nu H_2(x)
\end{equation}
with
\begin{equation}\label{ham}
H_0(x):=\frac{2\kappa+1}{4x(1-x)}\begin{pmatrix}
0 & 1\\
1 & 0
\end{pmatrix},\quad
H_1(x)=\begin{pmatrix}
2-\frac{1}{1-x} & 0\\
0 & 2-\frac{1}{x}
\end{pmatrix},\quad
H_2(x)=\begin{pmatrix}
0 & 2\\
2 & 0
\end{pmatrix}.
\end{equation}
At this point a comment is in order. We observe that the Hamiltonian system depends linearly on the physical parameters, the matrix $W$ is positive on the interval $(0,1)$ and both $W$ and $H$ are Hermitian. For the construction of a self-adjoint extension of the operator defined in (\ref{tau}) with $H$ given as in (\ref{CPAEHamil}) and (\ref{ham}) we refer to \cite{Batic-Schmid05}. Moreover, the differential expression $\tau_0$ introduced in (\ref{tau0}) where $H_0$ should be taken as in (\ref{ham}) is singular on $(0,1)$ and in the limit point case at both boundary points. Hence, the closure of the associated minimal operator $T_0=T_{min}$ is the only self-adjoint extension of $\tau_0$ in $L^2_W\left((0,1),\mathbb{C}^{2}\right)$. Finally, it was already proven in \cite{Batic-Schmid05} that the eigenvalues of $T_0$ are simple, isolated and real with algebraic multiplicity one. It remains to verify that it is possible to construct a differentiable matrix valued function $G:(0,1)\times\mathbb{C}\times\mathbb{C}^2\longrightarrow M_2(\mathbb{C})$ and maps $f_1,f_2,g:\mathbb{C}\times\mathbb{C}^2\longrightarrow\mathbb{C}$ satisfying the deformation equation (\ref{defeq}) with $m=2$, that is 
\begin{equation} \label{CPAEdeform}
    \frac{\partial G}{\partial x}+(\lambda W+H)JG-GJ(\lambda W+H)=f_1 H_1+f_2 H_2+gW
\end{equation}
with $H_1$ and $H_2$ given by (\ref{ham}). We will try the ansatz
\begin{equation}\label{G}
    G(x)=\begin{pmatrix}
        \widehat{a}x & \widehat{c}(1-x) \\
        \widehat{b}x & \widehat{d}(1-x)
    \end{pmatrix},
\end{equation}
where $\widehat{a}$, $\widehat{b}$, $\widehat{c}$ and $\widehat{d}$ are some unknown nonzero scalars. If we replace (\ref{G}) into (\ref{CPAEdeform}) and equate the entries of the matrices arising from the l.h.s. and r.h.s. of (\ref{CPAEdeform}), we end up with the following system of equations 
\begin{eqnarray}
4[\mu(\widehat{b}-\widehat{c})-2\widehat{a}\nu]x^2+2[4\widehat{a}\nu-\widehat{b}(\lambda+\mu)+\widehat{c}(\lambda+3\mu)+\widehat{a}-2f_1]x+\widehat{a}(2\kappa-1)-2\widehat{c}(\lambda+\mu)+2(f_1+g)&=&0,\label{aa1}\\
2\mu(\widehat{a}+\widehat{d})x+\widehat{a}(\lambda-\mu)-\widehat{d}(\lambda+\mu)+\widehat{c}+2f_2&=&0,\label{aa2}\\
2\mu(\widehat{a}+\widehat{d})x+\widehat{a}(\lambda-\mu)-\widehat{d}(\lambda+\mu)-\widehat{b}+2f_2&=&0,\label{aa3}\\
4[\mu(\widehat{c}-\widehat{b})-2\widehat{d}\nu]x^2+2[4\widehat{d}\nu-\widehat{b}(\lambda-\mu)+\widehat{c}(\lambda-3\mu)-\widehat{d}-2f_1]x+\widehat{d}(2\kappa+1)-2\widehat{c}(\lambda-\mu)+2(f_1-g)&=&0,\label{aa4}
\end{eqnarray}
which is satisfied for any $x\in(0,1)$ provided that the coefficients vanish. From (\ref{aa2}) and (\ref{aa3}) we immediately see that $\widehat{d}=-\widehat{a}$ and $\widehat{c}=-\widehat{b}$ and we end up with the underdetermined system of equations
\begin{eqnarray}
2\widehat{a}\lambda-\widehat{b}+2f_2&=&0,\\
\widehat{b}\mu -\widehat{a}\nu &=&0,\\
(1+4\nu)\widehat{a}-2\widehat{b}(\lambda+2\mu)-2f_1&=&0,\\
(1-4\nu)\widehat{a}-2\widehat{b}(\lambda-2\mu)-2f_1&=&0,\\
(2\kappa-1)\widehat{a}+2\widehat{b}(\lambda+\mu)+2(f_1+g)&=&0,\\
-(2\kappa+1)\widehat{a}+2\widehat{b}(\lambda-\mu)+2(f_1-g)&=&0,
\end{eqnarray}
admitting the solution
\begin{equation}
\widehat{a}=\widehat{a},\quad
\widehat{b}=\frac{\nu}{\mu}\widehat{a},\quad
f_1=\frac{\mu-2\lambda\nu}{2\mu}\widehat{a},\quad
f_2=\frac{\nu-2\lambda\mu}{2\mu}\widehat{a},\quad
g=-(\kappa+\nu)\widehat{a}.
\end{equation}
Finally, by means of (\ref{PDEeig}) we obtain the following PDE for the eigenvalues
\begin{equation}\label{PDE1}
(\mu-2\nu\lambda)\frac{\partial\lambda}{\partial\mu}+(\nu-2\mu\lambda)\frac{\partial\lambda}{\partial\nu}+2\mu(\kappa+\nu)=0,
\end{equation}
which agrees with equation (4) in  \cite{Batic-Schmid05} where the same equation was obtained by some fortuitous integrations by parts. It is interesting to observe, that if we apply the same approach to the special case $\mu=\nu$, it is possible to derive a very simple ODE for the eigenvalues. To this purpose, the deformation equation reads
\begin{equation} \label{new}
    \frac{\partial G}{\partial x}+(\lambda W+H)JG-GJ(\lambda W+H)=f_\mu\frac{\partial H}{\partial\mu}+gW
\end{equation}
with
\begin{equation}
H(x;\mu):=\begin{pmatrix}
        2\mu-\frac{\mu}{1-x} & 2\mu+\frac{2\kappa+1}{4x(1-x)} \\
        2\mu+\frac{2\kappa+1}{4x(1-x)}  & 2\mu-\frac{\mu}{x}
    \end{pmatrix}
\end{equation}
and $G$ given by the same ansatz we used before. A simple computation leads to
\begin{equation}
\widehat{a}=\widehat{b},\quad \widehat{b}=\widehat{b},\quad
f_\mu=\frac{1-2\lambda}{2}\widehat{b},\quad
g=-(k+\mu)\widehat{b}
\end{equation}
and hence, (\ref{PDEeig}) leads to the following ODE for the eigenvalues 
\begin{equation}
\frac{d\lambda}{d\mu}=\frac{2(\kappa+\mu)}{2\lambda-1}.
\end{equation}
The appropriate initial condition at $\mu=0$ is
\begin{equation}\label{ICl}
\lambda_{n,\kappa}(0)=\mbox{sgn}(n)\left(|\kappa|+|n|-\frac{1}{2}\right),\quad |\kappa|\geq\frac{1}{2},\quad\kappa=k-\frac{1}{2},\quad k\in\mathbb{Z},\quad n\in\mathbb{Z}\backslash\{0\}
\end{equation} 
See Appendix A in \cite{Batic-Schmid05} for more details. At this point, a straightforward integration gives
\begin{equation}\label{pip}
\lambda_{n,\kappa}(\mu)=\frac{1}{2}+\mbox{sgn}(n)\sqrt{\mu^2+2\kappa\mu+\left(\lambda_{n,\kappa}(0)-\frac{1}{2}\right)^2}.
\end{equation}
It is gratifying to see that (\ref{pip}) not only coincides with (45) in \cite{Batic-Schmid05} and a similar result in \cite{Suff} but its derivation through the deformation theory of linear Hamiltonian systems is much more efficient than the methods used in \cite{Suff,Batic-Schmid05}. Finally, we point out that it is not possible to obtain an ordinary differential equation where the independent variable is $\mu$ or $\nu$. To do that, we can introduce the deformation equations 
\begin{equation}\label{def1}
\frac{\partial G}{\partial x}+(\lambda W+H)JG-GJ(\lambda W+H)=f_\mu\frac{\partial H}{\partial\mu}+gW
\end{equation}
or
\begin{equation}\label{def2}
\frac{\partial G}{\partial x}+(\lambda W+H)JG-GJ(\lambda W+H)=f_\nu\frac{\partial H}{\partial\nu}+gW
\end{equation}
with
\begin{equation}
H(x;\mu,\nu):=\begin{pmatrix}
        2\mu-\frac{\mu}{1-x} & 2\nu+\frac{2\kappa+1}{4x(1-x)} \\
        2\nu+\frac{2\kappa+1}{4x(1-x)}  & 2\mu-\frac{\mu}{x}
    \end{pmatrix}
\end{equation}
and $G$ given as in (\ref{G}). It turns out that (\ref{def1}) and (\ref{def2}) are satisfied only in the trivial case $\widehat{a}=\widehat{b}=\widehat{c}=\widehat{d}=0$ and $f_\mu=f_\nu=0$. We conclude this section by observing that in the case the BH spin parameter $a$ and the energy particle $\omega$ are considered as independent variables, then by means of the ansatz (\ref{G}) and a deformation equation 
\begin{equation}
\frac{\partial G}{\partial x}+(\lambda W+H)JG-GJ(\lambda W+H)=f_a\frac{\partial H}{\partial a}+f_\omega\frac{\partial H}{\partial\omega}+gW
\end{equation}
with Hamiltonian
\begin{equation}
H(x;a,\omega):=\begin{pmatrix}
        2ma-\frac{ma}{1-x} & 2a\omega+\frac{2\kappa+1}{4x(1-x)} \\
        2a\omega+\frac{2\kappa+1}{4x(1-x)}  & 2ma-\frac{am}{x}
    \end{pmatrix}
\end{equation}
we can proceed as before to find
\begin{equation}
\widehat{a}=\widehat{a},\quad
b=\frac{\omega}{m}\widehat{a},\quad
f_a=-\frac{2\omega\lambda-m}{2m^2}\widehat{a},\quad
f_\omega=\frac{(\omega^2-m^2)\lambda}{m^2 a}\widehat{a},\quad
g=-(\kappa+a\omega)\widehat{a}
\end{equation}
and the corresponding PDE for the eigenvalues reads
\begin{equation}\label{PDEa}
a(2\omega\lambda-m)\frac{\partial\lambda}{\partial a}-2(\omega^2-m^2)\lambda\frac{\partial\lambda}{\partial\omega}=2m^2 a(\kappa+a\omega).
\end{equation}
with $|\omega|\geq m$ and $a\in\mathbb{R}$. We conclude this section with a remark. It would be tempting to use the method of characteristics to solve the corresponding Cauchy problems associated to the PDEs (\ref{PDE1}) and (\ref{PDEa}). Regarding the PDE (\ref{PDE1}) let $\mathfrak{a}(\mu,\nu,\lambda)=\mu-2\nu\lambda$, $\mathfrak{b}(\mu,\nu,\lambda)=\nu-2\mu\lambda$ and $\mathfrak{c}(\mu,\nu,\lambda)=2\mu(\kappa+\nu)$. In this case the initial curve is defined through the parametric equations $\mu(0,s)=x_0(s)=s$, $\nu(0,s)=y_0(s)=s$ and $\lambda(0,s)=\lambda_0(s)=\lambda_{n,\kappa}(s)$ given by (\ref{pip}) where $\mu$ is replaced by $s$. Even though the coefficients $\mathfrak{a}$, $\mathfrak{b}$ and $\mathfrak{c}$ are differentiable in  neighbourhood of the initial curve, the transversality condition allowing the change of variables $(\mu,\nu)$ to $(t,s)$, i.e.
\begin{equation}\label{trav}
J(s)=\mathfrak{a}(x_0(s),y_0(s),\lambda_0(s))\frac{dy_0}{ds}-\mathfrak{b}(x_0(s),y_0(s),\lambda_0(s))\frac{dx_0}{ds}\neq 0
\end{equation}
is clearly violated and therefore, the existence and uniqueness theorem for quasi-linear first order PDEs cannot be applied \cite{Rubi}. It is not difficult to verify that the same problem appears if we try to solve (\ref{PDEa}) along the initial curve $a(0,s)=0$,   $\omega(0,s)=s$ with $|s|\geq m$ and $\lambda(0,s)=\lambda_0(s)=\lambda_{n,\kappa}(0)$  given by (\ref{ICl}). It is interesting to get an idea of which kind of difficulty we may face in solving the problem of characteristics in the case we assign to (\ref{PDEa}) an initial curve satisfying (\ref{trav}). It can be readily checked that the characteristics are controlled by the following system of ODEs
\begin{eqnarray}
\frac{da}{dt}&=&a(2\omega\lambda-m),\label{ce1}\\
\frac{d\omega}{dt}&=&-2(\omega^2-m^2)\lambda,\label{ce2}\\
\frac{d\lambda}{dt}&=&2m^2 a(\kappa+a\omega).\label{ce3}
\end{eqnarray}
First of all, we can get rid of $\lambda$ if we multiply (\ref{ce1}) by $2m$ and add to it (\ref{ce2}). This leads to the equation
\begin{equation}
2m\frac{d}{dt}\left(\ln{a}+mt\right)+\omega\frac{d}{dt}\ln{\frac{\omega-m}{\omega+m}}=0
\end{equation}
which can be cast into the more compact form
\begin{equation}\label{ce5}
\frac{d}{dt}\ln{a^2(\omega^2-m^2)}=-2m.
\end{equation}
Moreover, from equation (\ref{ce2}) it follows that $\lambda$ is known and given by
\begin{equation}
\lambda(t)=-\frac{1}{2(\omega^2-m^2)}\frac{d\omega}{dt}
\end{equation}
whenever $\omega$ as a function of $t$ is known. In order to get another ODE involving only $a$ and $\omega$, we can differentiate (\ref{ce2}) with respect to $t$ and use (\ref{ce3}). We end up with 
\begin{equation}\label{omega}
\frac{d^2}{dt^2}\ln{\frac{\omega-m}{\omega+m}}=-8m^3 a(\kappa+a\omega).
\end{equation}
At this point, we can either look at the above equation as a quadratic equation in $a$ or we can directly integrate (\ref{ce5}), i.e.
\begin{equation}\label{kk}
a^2(\omega^2-m^2)=c_1 e^{-2mt},
\end{equation}
where the integration constant $c_1$ is positive because $\omega\geq m$. By means of (\ref{kk}) it is possible to express $a$ in terms of $\omega$ and replacing such an  expression into (\ref{omega}) gives rise to the following ODE
\begin{equation}
\frac{d^2\omega}{dt^2}-\frac{2\omega}{\omega^2-m^2}\left(\frac{d\omega}{dt}\right)^2=-4m^2\left(c_1\omega e^{-2mt}+\kappa\sqrt{c_1(\omega^2-m^2)}e^{-mt}\right).
\end{equation}
Hence, we showed that the characteristics system can be decoupled however solving the ODE for $\omega$ due to its nonlinear nature is not a trivial task. This is the reason why in the next section we decided to follow a different strategy in order to study the angular eigenvalues.

\section{Perturbative expansions of the eigenvalues}\label{EIGENODE}
In the following, we focus our attention to the PDE (\ref{PDEa}) and the derivation of its perturbative solutions. We start by considering the case $n>0$, $a\in\mathbb{R}$ and $\omega\geq m$ and construct a perturbative representation of the angular eigenvalues where we treat $a$ and $\epsilon=\omega-m$ as small parameters. As an approximation to the full solution $\lambda$, we consider a series like the following
\begin{equation}\label{expansion}
\lambda_{n,\kappa}(a,\omega)=\sum_{\substack{i,j=0\\ 0\leq i+j\leq 3}}^3\mathfrak{C}_{ij}a^i\epsilon^j+\mathcal{O}\left(a^\alpha\epsilon^\beta\right)
\end{equation}
with nonnegative integers $\alpha$ and $\beta$ such that $\alpha+\beta\geq 4$ and the starting condition
\begin{equation}
\mathfrak{C}_{00}=\lambda_{n,\kappa}(0)=|\kappa|+n-\frac{1}{2}
\end{equation}
which follows from (\ref{ICl}). If we replace (\ref{expansion}) into the PDE (\ref{PDEa}) and collect terms with
equal powers in $a$ and $\epsilon$, we end up with a system for the unknown coefficients $\mathfrak{C}_{ij}$ which we can be easily solved using Maple. More precisely, we find that 
\begin{eqnarray}
\mathfrak{C}_{01}&=&\mathfrak{C}_{02}=\mathfrak{C}_{12}=\mathfrak{C}_{03}=0,\\
\mathfrak{C}_{10}&=&\frac{m\kappa}{|\kappa|+n-1},\quad
\mathfrak{C}_{20}=\frac{m^2(n-1)(n-1+2|\kappa|)}{2(|\kappa|+n-1)^3},\quad
\mathfrak{C}_{11}=\frac{\kappa(2|\kappa|+2n-1)}{2(|\kappa|+n-1)(|\kappa|+n)},\\
\mathfrak{C}_{30}&=&-\frac{m^3\kappa(n-1)(n-1+2|\kappa|)}{2(|\kappa|+n-1)^5},\quad
\mathfrak{C}_{21}=\frac{m(n-1)(n-1+2|\kappa|)}{2(|\kappa|+n-1)^3}
\end{eqnarray}
and the angular eigenvalues admit the representation
\begin{equation}\label{per}
\lambda_{n,\kappa}(a,\omega)=\mathfrak{C}_{00}+\mathfrak{C}_{10}a+\mathfrak{C}_{20}a^2+
\mathfrak{C}_{11}a(\omega-m)+\mathfrak{C}_{30}a^3+\mathfrak{C}_{21}a^2(\omega-m)+\mathcal{O}\left(a^\alpha(\omega-m)^\beta\right),\quad \alpha+\beta\geq 4.
\end{equation}
It is gratifying to observe that differently as in \cite{Suff,Kalnins} the problem of dividing by numbers which may be zero does not occur in the coefficients appearing in the expansion (\ref{per}) because being $n\geq 1$ and $|\kappa|\geq 1/2$ ensures that $|\kappa|+n-1\geq 1/2$. In the case $n<0$ and $\omega\geq m$, we can use again the expansion (\ref{expansion}) and the initial condition appropriate to this case, i.e.
\begin{equation}
\widehat{\mathfrak{C}}_{00}=\frac{1}{2}-|\kappa|-|n|
\end{equation}
is determined by (\ref{ICl}). Proceeding as before we end up with the following new set of coefficients
\begin{eqnarray}
\widehat{\mathfrak{C}}_{01}&=&\widehat{\mathfrak{C}}_{02}=\widehat{\mathfrak{C}}_{12}=\widehat{\mathfrak{C}}_{03}=0,\\
\widehat{\mathfrak{C}}_{10}&=&-\frac{m\kappa}{|n|+|\kappa|},\quad
\widehat{\mathfrak{C}}_{20}=-\frac{m^2 |n|(|n|+2|\kappa|)}{2(|n|+|\kappa|)^3},\quad
\widehat{\mathfrak{C}}_{11}=\frac{\kappa(1-2|n|-2|\kappa|)}{2(|n|+|\kappa|)(|n|+|\kappa|-1)},\\
\widehat{\mathfrak{C}}_{30}&=&\frac{m^3\kappa |n|(|n|+2|\kappa|)}{2(|n|+|\kappa|)^5},\quad
\widehat{\mathfrak{C}}_{21}=-\frac{m|n|(|n|+2|\kappa|)}{2(|n|+|\kappa|)^3}.
\end{eqnarray}
Note that also in this case there is no division by a number which may be zero because $n\leq -1$ and $|\kappa|\geq 1/2$ imply that $|n|+|\kappa|-1\geq 1/2$ and a fortiori $|n|+|\kappa|\geq 3/2$. We consider now the case when $n>0$ and $\omega\leq-m$ and in particular, we treat the black hole spin and $\delta=\omega+m$ as small parameters. Moreover, we consider the expansion
\begin{equation}\label{expansion1}
\lambda_{n,\kappa}(a,\omega)=\sum_{\substack{i,j=0\\ 0\leq i+j\leq 3}}^3\mathfrak{B}_{ij}a^i\delta^j+\mathcal{O}\left(a^\alpha\delta^\beta\right),\quad\alpha,\beta\in\mathbb{N},\quad \alpha+\beta\geq 4
\end{equation}
and the starting condition
\begin{equation}
\mathfrak{B}_{00}=|\kappa|+n-\frac{1}{2}.
\end{equation}
It is not difficult to verify that the expansion coefficients are given as follows
\begin{eqnarray}
\mathfrak{B}_{01}&=&\mathfrak{B}_{02}=\mathfrak{B}_{12}=\mathfrak{B}_{03}=0,\\
\mathfrak{B}_{10}&=&-\frac{m\kappa}{|\kappa|+n},\quad
\mathfrak{B}_{20}=\frac{m^2 n(n+2|\kappa|)}{2(|\kappa|+n)^3},\quad
\mathfrak{B}_{11}=\frac{\kappa(2|\kappa|+2n-1)}{2(|\kappa|+n-1)(|\kappa|+n)},\\
\mathfrak{B}_{30}&=&\frac{m^3\kappa n(n+2|\kappa|)}{2(|\kappa|+n)^5},\quad
\mathfrak{B}_{21}=-\frac{mn(n+2|\kappa|)}{2(|\kappa|+n)^3}.
\end{eqnarray}
Hence, the angular eigenvalues admit the representation
\begin{equation}\label{per11}
\lambda_{n,\kappa}(a,\omega)=\mathfrak{B}_{00}+\mathfrak{B}_{10}a+\mathfrak{B}_{20}a^2+
\mathfrak{B}_{11}a(\omega+m)+\mathfrak{B}_{30}a^3+\mathfrak{B}_{21}a^2(\omega+m)+\mathcal{O}\left(a^\alpha(\omega+m)^\beta\right),\quad \alpha+\beta\geq 4.
\end{equation}
Finally, if $n<0$ and the starting condition is chosen to be 
\begin{equation}
\widehat{\mathfrak{B}}_{00}=\frac{1}{2}-|\kappa|-|n|,
\end{equation}
then the corresponding coefficients are
\begin{eqnarray}
\widehat{\mathfrak{B}}_{01}&=&\widehat{\mathfrak{B}}_{02}=\widehat{\mathfrak{B}}_{12}=\widehat{\mathfrak{B}}_{03}=0,\\
\widehat{\mathfrak{B}}_{10}&=&\frac{m\kappa}{|n|+|\kappa|-1},\quad
\widehat{\mathfrak{B}}_{20}=-\frac{m^2(|n|-1)(|n|-1+2|\kappa|)}{2(|n|+|\kappa|-1)^3},\quad
\widehat{\mathfrak{B}}_{11}=\frac{\kappa(1-2|n|-2|\kappa|)}{2(|n|+|\kappa|)(|n|+|\kappa|-1)},\\
\widehat{\mathfrak{B}}_{30}&=&-\frac{m^3\kappa (|n|-1)(|n|-1+2|\kappa|)}{2(|n|+|\kappa|-1)^5},\quad
\widehat{\mathfrak{B}}_{21}=\frac{m(|n|-1)(|n|-1+2|\kappa|)}{2(|n|+|\kappa|-1)^3}.
\end{eqnarray}
We conclude this section with the observation that the PDE (\ref{PDEa}) can be also used in order to obtain a perturbative expansion for the angular eigenvalues in the case of a fast spinning Kerr naked singularity. To this purpose, we introduce a stretched spinning scale $\widetilde{a}=a/\gamma$ with $a$ fixed. In the limit of $\gamma\to 0$, $\widetilde{a}$ becomes very large and we can consider the case of a fast spinning naked singularity by requiring that $a^2>M^2$. The transformed version of (\ref{PDEa}) becomes
\begin{equation}
\widetilde{a}(2\omega\lambda-m)\frac{\partial\lambda}{\partial\widetilde{a}}-2(\omega^2-m^2)\lambda\frac{\partial\lambda}{\partial\omega}=2m^2\widetilde{a}\left(\kappa\gamma+\widetilde{a}\omega\gamma^2\right).
\end{equation}
If we introduce the perturbative expansion
\begin{equation}
\lambda(\widetilde{a},\omega)=\lambda_0(\widetilde{a},\omega)+\gamma\lambda_1(\widetilde{a},\omega)+\mathcal{O}(\gamma^2),
\end{equation}
we find that $\lambda_0$ and $\lambda_1$ must satisfy the PDEs
\begin{eqnarray}
\widetilde{a}(2\omega\lambda_0-m)\frac{\partial\lambda_0}{\partial\widetilde{a}}-2(\omega^2-m^2)\lambda_0\frac{\partial\lambda_0}{\partial\omega}&=&0,\label{ek1}\\
\widetilde{a}(2\omega\lambda_0-m)\frac{\partial\lambda_1}{\partial\widetilde{a}}-2(\omega^2-m^2)\lambda_0\frac{\partial\lambda_1}{\partial\omega}+2\left[\widetilde{a}\omega\frac{\partial\lambda_0}{\partial\widetilde{a}}-(\omega^2-m^2)\frac{\partial\lambda_0}{\partial\omega}\right]\lambda_1&=&2m^2\widetilde{a}\kappa.\label{ek2}
\end{eqnarray}
Equation (\ref{ek1}) can be solved with Maple and a particular solution in implicit form is found to be 
\begin{equation}
\widetilde{a}-(\omega-m)^{-\frac{2\lambda_0-1}{4\lambda_0}}(\omega+m)^{-\frac{2\lambda_0+1}{4\lambda_0}}=0.
\end{equation}
Solving the above equation for $\lambda_0$ yields after some straightforward algebra the following expression for $\lambda_0$, namely
\begin{equation}\label{l0}
\lambda_0(\widetilde{a},\omega)=\frac{1}{\ln{\widetilde{a}^2(\omega^2-m^2)}}\ln{\sqrt{\frac{\omega-m}{\omega+m}}}.
\end{equation}
Even though (\ref{ek2}) is a non-homogeneous linear first order PDE, after replacing (\ref{l0}) into (\ref{ek2}) the coefficients of the corresponding PDE becomes complicated functions of $\widetilde{a}$ and $\omega$ and any attempt to construct a particular solution results in a daunting task.

\section{Conclusions and outlook}\label{Conclusions}
In this paper we have applied a newly result in the deformation theory of linear Hamiltonian systems to the angular eigenvalue problem for the Dirac equation in the Kerr metric. We derived an explicit expression for the deformation matrix leading to a first order quasi-linear PDE for the angular eigenvalues. Such a PDE has been directly expressed in terms of the relevant physical parameters of the problem, i.e. the angular momentum per unit mass of the BH and the particle energy. We proved that it is not possible for the eigenvalues to satisfy an ODE with respect to one of the physical parameters as previously claimed in the literature. Moreover, we were able to show that the Cauchy problem for the aforementioned PDE is ill-defined for the initial data considered in the present work. In this context, by a non-trivial decoupling of the characteristic system we pointed out which overwhelming difficulties may arise even in the case of choices of the initial curve leading to a well-defined Cauchy problem. For this reason, we decided to adopt a different strategy in the analysis of the angular eigenvalues. More precisely, we constructed a perturbative solution of the PDE such that in the limit of a vanishing BH spin parameter it correctly reproduces the angular eigenvalues of the Dirac equation in the Schwarzschild metric. Finally, by a scaling technique we constructed a perturbative expansion for the eigenvalues in the case of a fast rotating naked Kerr singularity. In a forthcoming work, we will consider applications of the result obtained by \cite{Schmid21} in several general relativistic contexts.

\bigskip

{\bf{Data accessibility}} This article does not use data.

\end{document}